\documentclass[aps,prc,twocolumn,showpacs,preprintnumbers,superscriptaddress]{revtex4}
\usepackage{graphicx,epsfig}

\newcommand{ \be }{\begin{equation}}
\newcommand{ \ee }{\end{equation}}
\newcommand{ \bea }{\begin{eqnarray}}
\newcommand{ \eea }{\end{eqnarray}}

\usepackage{color}

\begin{document}

\title{Global polarization measurement in Au+Au collisions}

\affiliation{Argonne National Laboratory, Argonne, Illinois 60439}
\affiliation{University of Birmingham, Birmingham, United Kingdom}
\affiliation{Brookhaven National Laboratory, Upton, New York 11973}
\affiliation{California Institute of Technology, Pasadena, California 91125}
\affiliation{University of California, Berkeley, California 94720}
\affiliation{University of California, Davis, California 95616}
\affiliation{University of California, Los Angeles, California 90095}
\affiliation{Carnegie Mellon University, Pittsburgh, Pennsylvania 15213}
\affiliation{University of Illinois at Chicago, Chicago, Illinois 60607}
\affiliation{Creighton University, Omaha, Nebraska 68178}
\affiliation{Nuclear Physics Institute AS CR, 250 68 \v{R}e\v{z}/Prague, Czech Republic}
\affiliation{Laboratory for High Energy (JINR), Dubna, Russia}
\affiliation{Particle Physics Laboratory (JINR), Dubna, Russia}
\affiliation{University of Frankfurt, Frankfurt, Germany}
\affiliation{Institute of Physics, Bhubaneswar 751005, India}
\affiliation{Indian Institute of Technology, Mumbai, India}
\affiliation{Indiana University, Bloomington, Indiana 47408}
\affiliation{Institut de Recherches Subatomiques, Strasbourg, France}
\affiliation{University of Jammu, Jammu 180001, India}
\affiliation{Kent State University, Kent, Ohio 44242}
\affiliation{Institute of Modern Physics, Lanzhou, China}
\affiliation{Lawrence Berkeley National Laboratory, Berkeley, California 94720}
\affiliation{Massachusetts Institute of Technology, Cambridge, MA 02139-4307}
\affiliation{Max-Planck-Institut f\"ur Physik, Munich, Germany}
\affiliation{Michigan State University, East Lansing, Michigan 48824}
\affiliation{Moscow Engineering Physics Institute, Moscow Russia}
\affiliation{City College of New York, New York City, New York 10031}
\affiliation{NIKHEF and Utrecht University, Amsterdam, The Netherlands}
\affiliation{Ohio State University, Columbus, Ohio 43210}
\affiliation{Panjab University, Chandigarh 160014, India}
\affiliation{Pennsylvania State University, University Park, Pennsylvania 16802}
\affiliation{Institute of High Energy Physics, Protvino, Russia}
\affiliation{Purdue University, West Lafayette, Indiana 47907}
\affiliation{Pusan National University, Pusan, Republic of Korea}
\affiliation{University of Rajasthan, Jaipur 302004, India}
\affiliation{Rice University, Houston, Texas 77251}
\affiliation{Universidade de Sao Paulo, Sao Paulo, Brazil}
\affiliation{University of Science \& Technology of China, Hefei 230026, China}
\affiliation{Shanghai Institute of Applied Physics, Shanghai 201800, China}
\affiliation{SUBATECH, Nantes, France}
\affiliation{Texas A\&M University, College Station, Texas 77843}
\affiliation{University of Texas, Austin, Texas 78712}
\affiliation{Tsinghua University, Beijing 100084, China}
\affiliation{Valparaiso University, Valparaiso, Indiana 46383}
\affiliation{Variable Energy Cyclotron Centre, Kolkata 700064, India}
\affiliation{Warsaw University of Technology, Warsaw, Poland}
\affiliation{University of Washington, Seattle, Washington 98195}
\affiliation{Wayne State University, Detroit, Michigan 48201}
\affiliation{Institute of Particle Physics, CCNU (HZNU), Wuhan 430079, China}
\affiliation{Yale University, New Haven, Connecticut 06520}
\affiliation{University of Zagreb, Zagreb, HR-10002, Croatia}

\author{B.I.~Abelev}\affiliation{University of Illinois at Chicago, Chicago, Illinois 60607}
\author{M.M.~Aggarwal}\affiliation{Panjab University, Chandigarh 160014, India}
\author{Z.~Ahammed}\affiliation{Variable Energy Cyclotron Centre, Kolkata 700064, India}
\author{B.D.~Anderson}\affiliation{Kent State University, Kent, Ohio 44242}
\author{D.~Arkhipkin}\affiliation{Particle Physics Laboratory (JINR), Dubna, Russia}
\author{G.S.~Averichev}\affiliation{Laboratory for High Energy (JINR), Dubna, Russia}
\author{Y.~Bai}\affiliation{NIKHEF and Utrecht University, Amsterdam, The Netherlands}
\author{J.~Balewski}\affiliation{Indiana University, Bloomington, Indiana 47408}
\author{O.~Barannikova}\affiliation{University of Illinois at Chicago, Chicago, Illinois 60607}
\author{L.S.~Barnby}\affiliation{University of Birmingham, Birmingham, United Kingdom}
% ...

\collaboration{STAR Collaboration}\noaffiliation

\date{\today}

\begin{abstract}
The system created in non-central relativistic nucleus-nucleus collisions possesses large
orbital angular  momentum. Due to spin-orbit coupling, particles produced in such a system
could become globally polarized along the direction of the system angular momentum. We
present the results of $\Lambda$ and $\bar\Lambda$ hyperon global polarization
measurements in Au+Au collisions at $\sqrt{s_{NN}}=62.4$~GeV and $200$~GeV performed with the
STAR detector at RHIC. The observed global polarization of $\Lambda$ and $\bar\Lambda$
hyperons in the STAR acceptance is consistent with zero within the precision of the
measurements. The obtained upper limit,  $|P_{\Lambda,\bar\Lambda}| \leq 0.02$, is
compared to the theoretical values discussed recently in the literature.
\end{abstract}


\maketitle

\section{Introduction}
\label{Introduction} The system created in non-central relativistic nucleus-nucleus
collisions possesses large orbital angular momentum.
One of the novel phenomena predicted to occur in such a system is global system
polarization~\cite{Liang:2004ph,Voloshin:2004ha,Liang:2004xn}. This phenomenon manifests
itself in the polarization of secondary produced particles along the direction of the
system angular momentum. The global polarization may provide valuable insights into the
evolution of the system, the hadronization mechanism, and the origin of hadronic spin
preferences.  The system orbital angular momentum may be transformed into global particle
spin orientation preferences by spin-orbit coupling at various stages of the system
evolution. It can happen at the partonic level, while the system evolves as an ensemble of
deconfined polarized quarks. The polarization of the secondary produced hadrons could also
be acquired via hadron re-scattering at a later hadronic stage. An example of such system
orbital momentum transformation into global polarization of produced $\rho$-mesons, due to
pion re-scattering, is discussed in~\cite{Voloshin:2004ha}.

One specific scenario for the spin-orbit transformation via the polarized quark phase is
discussed in~\cite{Liang:2004ph}. There, it is argued that parton interactions in
non-central relativistic nucleus-nucleus collisions lead first to the global polarization
of the produced quarks. The values for this global quark polarization at RHIC
(Relativistic Heavy Ion Collider) energies were estimated to be quite high, around 30\%
percent. In the case of a strongly interacting QGP (Quark Gluon Plasma), this global quark
polarization can have many observable consequences, such as a left-right asymmetry in
hadron production at large rapidity (similar to the single-spin asymmetry in $pp$
collisions) or polarization of thermal photons, di-leptons, and final hadrons with
non-zero spin. In particular, it would lead to global polarization of the hyperons, which
could be measured via their weak, self-analyzing decays. Assuming that the strange and
non-strange quark polarizations, $P_s$ and $P_q$, are equal, in the particular case of the
`exclusive' parton recombination scenario~\cite{Liang:2004ph}, the values of the global
polarization $P_H$ for $\Lambda$, $\Sigma$, and $\Xi$ hyperons appear to be similar to
those for quarks: $P_H = P_q \simeq 0.3$. Recently more realistic calculations
\cite{Liang:2007ma} of the global quark polarization were performed within a model based
on the HTL (Hard Thermal Loop) gluon propagator. The resulting hyperon polarization was
predicted to be in the range from $-0.03$ to $0.15$ depending on the temperature of the
QGP formed.

In this paper we present the results of $\Lambda$ and $\bar\Lambda$ hyperon global
polarization measurements in Au+Au collisions performed at $\sqrt{s_{NN}}$=62.4 and
200~GeV with the STAR (Solenoidal Tracker At RHIC) detector. In this work the
polarization is defined to be positive if the hyperon spin has a positive component along
the system orbital momentum, while in~\cite{Liang:2004ph,Liang:2007ma} the opposite
convention is used. The paper is organized as follows. First we overview the global
polarization measurement technique and introduce relevant observables. Then the results of
$\Lambda$ and $\bar\Lambda$ hyperon global polarization are presented as functions of
pseudo-rapidity, transverse momentum, and collision centrality. Subsequently, the possible
systematic uncertainties of the method and the detector acceptance effects are discussed
and systematic errors are estimated.

\section{Global polarization of hyperons}
\label{GlobalPolarization}

The global polarization of hyperons can be determined from the angular
distribution of hyperon decay products relative to the system orbital
momentum {\boldmath $L$}:
\begin{eqnarray}
\label{GlobalPolarizationDefinition}
\frac{dN}{d \cos \theta^*} \sim 1+\alpha_H~P_H~\cos \theta^* ~,
\end{eqnarray}
where $P_H$ is the hyperon global polarization, $\alpha_H$ is the hyperon decay parameter,
and $\theta^*$ is the angle in the hyperon rest frame between the system orbital momentum
{\boldmath $L$} and the 3-momentum of the baryon daughter from the hyperon decay.

The global polarization $P_H$ in Eq.~\ref{GlobalPolarizationDefinition} can depend on
hyperon kinematic variables such as transverse momentum $p_t^H$
and pseudorapidity $\eta^H$, as well as on the
relative azimuthal angle between the hyperon 3-momentum
and the direction of the system orbital momentum {\boldmath $L$}.
In this work we report the $p_t^H$ and $\eta^H$ dependence of the global polarization
averaged over the relative azimuthal angle
(see section~\ref{SystematicsUncertainties} for a detailed discussion of this definition).

Since the system angular momentum {\boldmath $L$}  is perpendicular to the reaction plane,
the global polarization can be measured via the distribution of the azimuthal angle of the
hyperon decay baryon (in the hyperon rest frame) with respect to the reaction plane.
Thus, the known and well established anisotropic flow measurement
techniques~\cite{Voloshin:1994mz,Poskanzer:1998yz} can be applied.

\begin{figure}[th]
\includegraphics[width=0.5\textwidth]{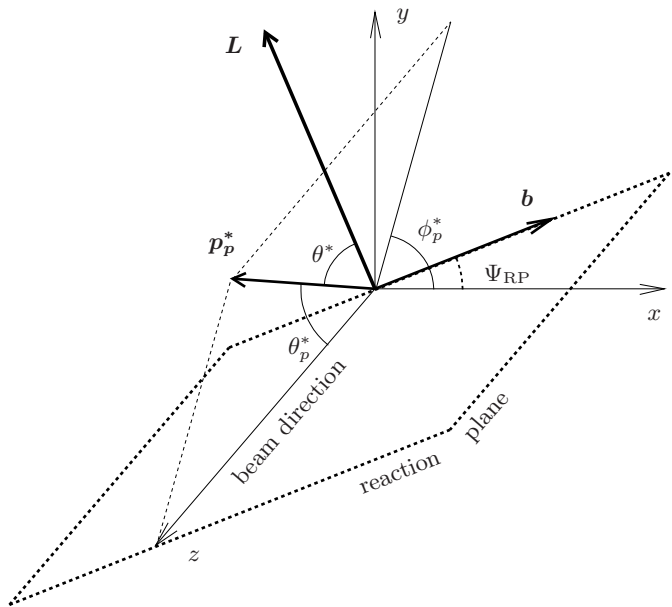}
\put(-172,224){{\boldmath $L$}} \put(-186,32){$z$}
\put(-170,60){\rotatebox{49}{beam~direction}} \put(-107,237){$y$} \put(-12,123){$x$}
\put(-60,166){{{\boldmath $b$}}} \put(-74,138){$\Psi_{\rm RP}$} \put(-178,150){\boldmath
$p^*_p$} \put(-99,155){$\phi^*_p$} \put(-148,110){$\theta^*_p$} \put(-139,145){$\theta^*$}
\put(-122,57){\rotatebox{22}{reaction}} \put(-83,77){\rotatebox{49}{plane}}
\caption{
\label{notaionsPlot}
Diagram showing the notations for the different angles
adopted in this paper. The laboratory frame is defined by the $x$, $y$, and $z$ (beam
direction) axes. {\boldmath $p^*_p$} is the hyperon decay baryon 3-momentum in the
hyperon rest frame. The reaction plane is spanned by the impact parameter {\boldmath $b$}
and the beam direction. The normal to the reaction plane defines the direction of the
system orbital momentum {\boldmath $L$}. Reversal of the orbital momentum,
{\boldmath $L$}$\to$ -{\boldmath $L$}, corresponds to changing the reaction plane angle by
$\Psi_{\rm RP}\to \Psi_{\rm RP}+\pi$.
}
\end{figure}

In order to write an equation for the global polarization in terms of the
observables used in anisotropic flow measurements,
we start with the equation that directly follows from the global
polarization definition~(\ref{GlobalPolarizationDefinition}):
\be
\label{GlobalPolarizationMeanCos}
P_{H}~=~\frac{3}{\alpha_H}~\langle \cos \theta^*\rangle~~.
\ee
The angle brackets in this equation denote averaging over the solid angle of the hyperon
decay baryon 3-momentum in the hyperon rest frame and over all directions of the system
orbital momentum {\boldmath $L$}, or, in other words, over all possible orientations of
the reaction plane.
Similarly, we can write an equation for the global polarization in
terms of the reaction plane angle $\Psi_{\rm RP}$ and the azimuthal angle $\phi^*_p$ of the
hyperon decay baryon 3-momentum in the hyperon's rest frame (see Fig.~\ref{notaionsPlot}
for notations).
By using a trigonometric relation among the angles, $\cos \theta^* = \sin
\theta^*_p \cdot \sin \left( \phi^*_p - \Psi_{\rm RP}\right)$ ($\theta^*_p$ is the angle
between the hyperon's decay baryon 3-momentum in the hyperon rest frame and the beam
direction), and integrating distribution~(\ref{GlobalPolarizationDefinition}) over the
angle $\theta^*_p$, one finds the following equation for the global polarization:
\begin{eqnarray}
\label{GlobalPolarizationObservable}
P_{H}~=~\frac{8}{\pi\alpha_H}\langle \sin \left( \phi^*_p
- \Psi_{\rm RP}\right)\rangle~~.
\end{eqnarray}
In this equation, perfect detector acceptance is assumed.
See section~\ref{SystematicsUncertainties} for the discussion of the detector acceptance effects.

Equation~\ref{GlobalPolarizationObservable} is similar to that used in directed flow
measurements~\cite{Barrette:1996rs,Alt:2003ab,Adams:2005aa,Adams:2005ca}.
For example, the hyperon directed flow can be defined as $v_1^H =
\langle \cos \left( \phi_H - \Psi_{\rm RP}\right)\rangle$, where $\phi_H$ is the azimuthal
angle of the hyperon's transverse momentum. The similarity to
Eq.~\ref{GlobalPolarizationObservable} allows us to use the corresponding anisotropic flow
measurement technique, and in this paper we will follow the same naming conventions and
notations as those adopted in an anisotropic flow analysis.

\subsection{Technique}
\label{Technique}

The main components of the detector system used in this analysis are the STAR main TPC
(Time Projection Chamber) \cite{Anderson:2003ur}, two STAR Forward TPCs
\cite{Ackermann:2002yx} and the STAR ZDC SMD (Zero Degree Calorimeter Shower Maximum
Detector) \cite{Adler:2000bd,Allgower:2002zy,ZDCSMDproposal:2003}. Data taken with a
minimum-bias trigger have been used for this analysis. The collision centrality was
defined using the total charged particle multiplicity within a pseudorapidity window of
$|\eta|<0.5$. The charged particle multiplicity distribution was divided into nine
centrality bins (classes): {\mbox{0-5\%} (most central collisions)}, \mbox{5-10\%}, \mbox{10-20\%},
\mbox{20-30\%}, \mbox{30-40\%}, \mbox{40-50\%}, \mbox{50-60\%}, \mbox{60-70\%}, and \mbox{70-80\%} of the total hadronic
inelastic cross section for Au+Au collisions. Our analysis was restricted to events with a
primary vertex within 30~cm of the center of the TPC along the beam direction. This
yielded a data set of $8.3 \times 10^6$ ($9.1 \times 10^6$) minimum-bias events for Au+Au
collisions at $62.4$~GeV ($200$~GeV) recorded with the STAR detector during RHIC run IV
(year 2004).

The hyperon reconstruction procedure used in this analysis is similar to that in
\cite{Adler:2002pb,Cai:2005ph,Takahashi:2005pq}. The $\Lambda$ and $\bar\Lambda$ particles
were reconstructed from their weak decay topology, $\Lambda \to p \pi^- $ and $\bar\Lambda
\to \bar p \pi^+ $, using charged tracks measured in the TPC. The corresponding decay
parameter is $\alpha_{\Lambda}^{-} = - \alpha_{\bar\Lambda}^{+} = 0.642\pm0.013$
\cite{Eidelman:2004wy}. Particle assignments for $p$ ($\bar p$) and $\pi^-$ ($\pi^+$)
candidates were based on charge sign and the mean energy loss, $dE/dx$, measured for each
track with at least 15 recorded space hits in the TPC. Candidate tracks were then paired
to form neutral decay vertices, which were required to be at least 6~cm from
the primary vertex. The reconstructed momentum vector at the decay vertex was required to
point back to the primary event vertex within 0.5~cm. For the $\Lambda$ and $\bar\Lambda$
reconstruction we chose pion candidates with a dca (distance of closest approach) to the
primary vertex of more than 2.5~cm and proton candidates with a dca~$>1.0$~cm.

\begin{figure}[th]
\includegraphics[width=0.45\textwidth]{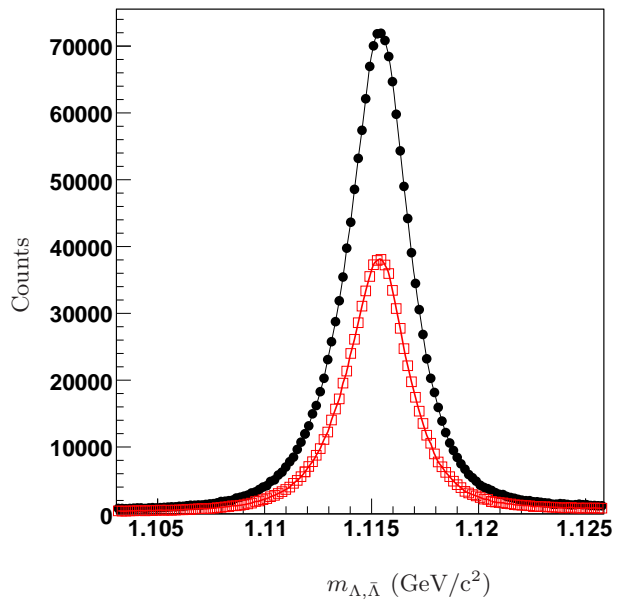}
\put(-240,100){\rotatebox{90}{Counts}} \put(-121,-5){$m_{\Lambda, \bar\Lambda}$ (GeV/c$^2$)}
\caption{
\label{lambdaInvMass}
(Color online) Invariant mass distribution for the
$\Lambda$ (filled circles) and $\bar\Lambda$ (open squares) candidates after the
quality cuts for Au+Au collisions at $\sqrt{s_{NN}}$=62.4~GeV  (centrality region \mbox{0-80\%}).}
\end{figure}

Figure~\ref{lambdaInvMass} shows the invariant mass distributions for the reconstructed
$\Lambda$ (filled circles) and $\bar\Lambda$ (open squares) candidates in the $|\eta_{\Lambda,
\bar\Lambda}|<1.3$ and $p_t^{\Lambda, \bar\Lambda}<4.5$ GeV/c region from the data sample
for Au+Au collisions at $62.4$~GeV. In this analysis the hyperon candidates with invariant
mass within the window $1.11<m_{\Lambda, \bar\Lambda}<1.12$ GeV/c$^2$ are used.
The background contribution, including $K^0_S$ meson contamination,
is estimated by fitting the invariant mass distribution with the sum of a Gaussian and 3rd-order polynomial function,
and is less than 8\%; it has been included in the estimate of the total
systematic errors.

The $\Lambda$ and $\bar\Lambda$ global polarization is calculated on the basis of
Eq.~\ref{GlobalPolarizationObservable}. The measured hyperons consist of primordial
$\Lambda$ ($\bar\Lambda$) and feed-downs from multistrange hyperons
($\Xi^0$ and $\Omega$) and $\Sigma^0$ decays, and also from short-lived resonances decaying via strong interactions.
The effect of these feed-downs, estimated as described below, is
incorporated in our systematic errors in subsection~\ref{SystematicsUncertainties}.
Under the assumption that the global polarization has the same value for
$\Lambda$ and $\Sigma^0$ \cite{Liang:2004ph}, we estimate the relative contribution from
$\Sigma^0$ to the extracted global polarization of the $\Lambda$ hyperons to be $\leq
30$\%.  This estimate takes into account an average polarization transfer from $\Sigma^0$
to $\Lambda$ of $-1/3$ \cite{Sucher:1965,Armenteros:1970eg} (this value can be
affected by non-uniform acceptance of the daughter $\Lambda$).
The $\Sigma^0/\Lambda$ production ratio is measured in d+Au collisions
at $\sqrt{s_{NN}}=200$~GeV to be 15\% \cite{VanBuren:2005px}
and is typically expected to be 2-3 times higher in Au+Au collisions.
Based on the results in \cite{Adams:2006ke}, the contribution of feed-downs from multiply strange hyperons ($\Xi$,
$\Omega$) is estimated to be less than 15\%. This can dilute the measured polarization and introduce
a similar systematic uncertainty ($\sim 15$\%) to the global polarization measurement.
The effect of feed-downs to $\Lambda$
($\bar\Lambda$) from strongly decaying resonances has not been measured with the STAR
detector. String fragmentation model calculations \cite{Pei:1997xt},
and study within the scenario of hadron gas fireball formation at thermal
and partial chemical equilibrium \cite{Becattini:1997rv},
suggest that in $pp$
collisions the fraction of direct hyperons is about 25-30\% for $\Lambda$ and 15-30\% for $\bar\Lambda$.

The global polarization measurement could also conceivably be affected by hyperon spin
precession in the strong magnetic field within the TPC. Using the equation for the spin
precession frequency, $\omega_H =2\mu_H B/\hbar$, one can estimate the shift of the
$\Lambda$ and $\bar\Lambda$ azimuthal spin orientation in the TPC magnetic field
($B=0.5$~T) at $p_{\Lambda,\bar\Lambda}=3.0$ GeV/c to be $|\delta\phi_{\Lambda,\bar\Lambda}|
\sim |\omega_{\Lambda,\bar\Lambda} * \tau_{\Lambda,\bar\Lambda} *
\gamma_{\Lambda,\bar\Lambda}| \sim 0.022$
($\gamma_{\Lambda,\bar\Lambda}$ is the hyperon's Lorentz factor).
For the hyperon magnetic moment $\mu_H$ and
mean lifetime $\tau_H$ we use values \cite{Eidelman:2004wy}: $\mu_{\Lambda,\bar\Lambda}= -
0.613$~$\mu_N$ (where $\mu_N$ is the nuclear magneton) and $\tau_{\Lambda,\bar\Lambda}=
2.63\times10^{-10}$~s. Thus, the effect of the spin precession on the global polarization
measurements is negligible ($\le 0.1$\%).

The reaction plane angle in Eq.~\ref{GlobalPolarizationObservable} is estimated by
calculating the so-called event plane flow vector $Q_{\rm EP}$. This implies the necessity to
correct the final results by the reaction plane resolution
$R_{\rm EP}$~\cite{Barrette:1996rs,Poskanzer:1998yz}. Similar to the case of directed flow,
the global polarization measurement requires knowledge of the direction of the system
orbital momentum {\boldmath $L$}, hence, of the first-order event plane vector. Taking
this into account, one can rewrite Eq.~\ref{GlobalPolarizationObservable} in terms of the
first-order event plane angle $\Psi_{\rm EP}^{(1)}$ and its resolution $R_{\rm EP}^{(1)}$:
\begin{eqnarray}
\label{GlobalPolarizationScalarProduct}
P_{H}~=~\frac{8}{\pi\alpha_H}\frac{\langle \sin \left(
  \phi^*_p - \Psi_{\rm EP}^{(1)}\right)\rangle}{R_{\rm EP}^{(1)}}~~.
\end{eqnarray}
There are a few different possibilities to determine the first-order event plane vector in
the STAR detector, using either the TPC, the Forward TPCs, or the ZDC SMD.
In this analysis,
the first-order event plane vector was determined from the Forward TPCs, which span a
pseudorapidity region ($2.7 < |\eta| < 3.9$) characterized by much larger directed flow
than the TPC region ($|\eta| < 1.3$). The charged particle tracks with transverse momentum
$0.15 < p_t < 2.0$ GeV/c are used to define the event plane vector.
The combination of two Forward TPC event plane vectors provides the full event plane.
The corresponding event-plane resolution, $R_{\rm EP}^{(1)}\{{\rm FTPC}\}$, is obtained from the
correlation of the two event plane vectors defined for two random subevents \cite{Voloshin:1994mz,Poskanzer:1998yz}.
Information on the
second-order event plane vector determined by the strong \emph{elliptic} flow in the TPC
pseudorapidity region was also used in this analysis, to calculate the systematic errors
coming from the uncertainty in the reaction plane definition. Use of the ZDC SMD to
determine the first-order reaction plane would introduce smaller systematic uncertainties,
but significantly poorer reaction plane resolution, compared to the use of the Forward
TPCs, and was not practical due to limited statistics. For more discussion on systematic
uncertainties and the role of the reaction plane resolution, see
subsection~\ref{SystematicsUncertainties}.

The direction of the system angular momentum in
Eq.~\ref{GlobalPolarizationScalarProduct} is fixed
by a convention, that spectator neutrons are
deflected along the direction of
the impact parameter and thus their
directed flow measured with ZDC SMD is positive for a positive pseudorapidity value.
From correlations between Forward TPC and ZDC SMD \cite{Adams:2005ca}
it follows that directed flow in the Forward TPC pseudorapidity region,
which is used to determine the first order event plane in this analysis,
has an opposite sign compared to that of spectator neutrons.
This is further taken into account when determining the direction of the system angular momentum.

\subsection{Results}
\label{UpperLimit}

\begin{figure}[th]
\includegraphics[width=0.5\textwidth]{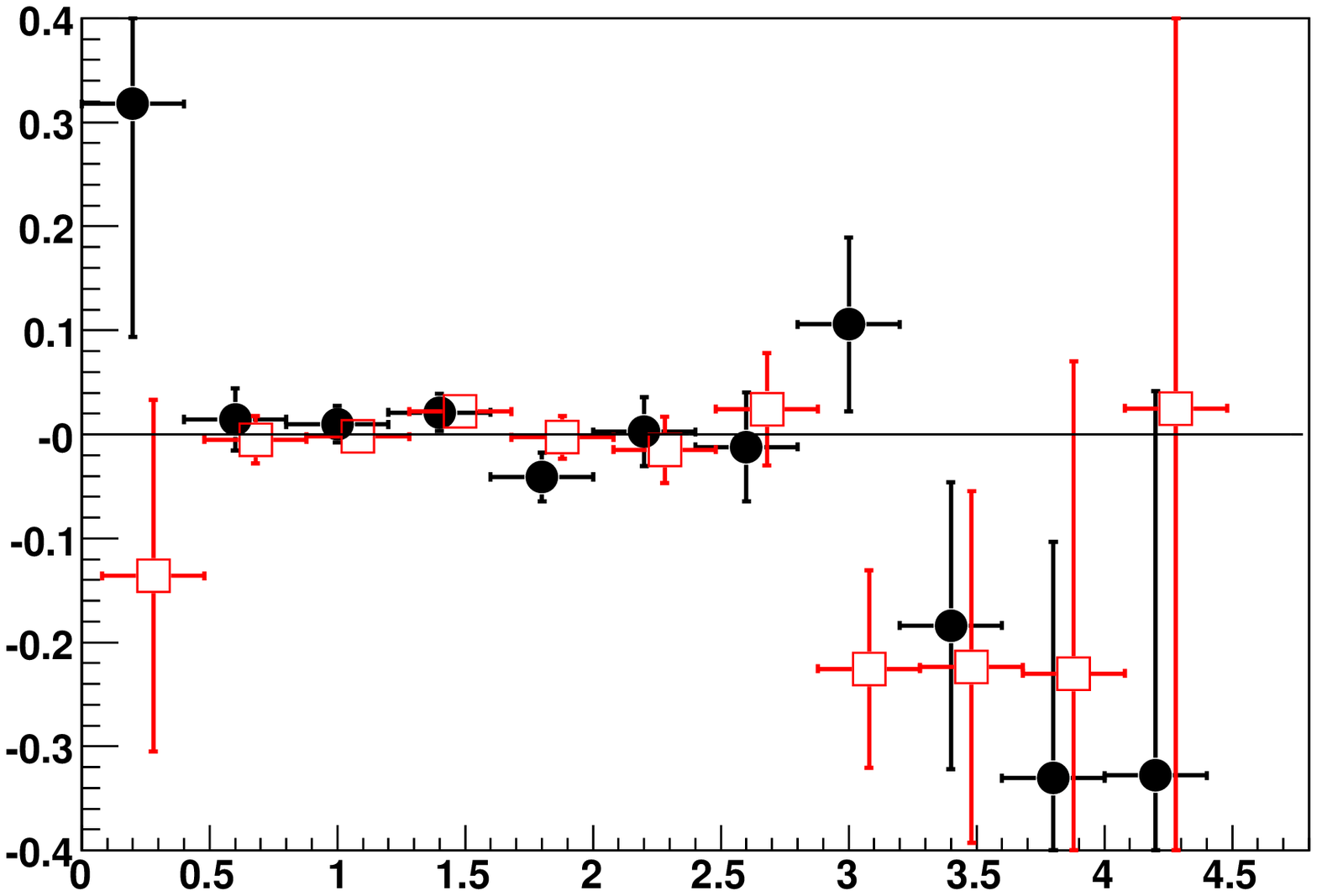}
\put(-256,82){\rotatebox{90}{$P_{\Lambda}$}}
\put(-131,-5){$p_t^{\Lambda}$ (GeV/c)}
\caption{
\label{lambdaGlobalPolarization_pt} (Color online) Global polarization of
$\Lambda$--hyperons as a function of $\Lambda$ transverse momentum $p_t^\Lambda$.
Filled circles show the results for Au+Au collisions at $\sqrt{s_{NN}}$=200~GeV
(centrality region \mbox{20-70\%}) and open squares indicate the results for Au+Au collisions at
$\sqrt{s_{NN}}$=62.4~GeV (centrality region \mbox{0-80\%}).
Only statistical uncertainties are shown.
}
\end{figure}

Figure~\ref{lambdaGlobalPolarization_pt} presents the $\Lambda$--hyperon global
polarization as a function of $\Lambda$ transverse momentum $p_t^{\Lambda}$
calculated on the basis of Eq.~\ref{GlobalPolarizationScalarProduct}.
The filled circles show the results of the measurement for Au+Au collisions at
$\sqrt{s_{NN}}$=200~GeV. The open squares indicate the results of a similar measurement for
Au+Au collisions at $\sqrt{s_{NN}}$=62.4~GeV.
The data points are corrected for the effects of the non-uniform detector acceptance.
Details on acceptance effects and systematic uncertainties are discussed in Sec.\ref{SystematicsUncertainties}.
Although the error bars at higher $\Lambda$
transverse momentum are rather large, there could be an indication in
Fig.~\ref{lambdaGlobalPolarization_pt} of a possible $p_t^\Lambda$ dependence of the
global polarization
(a constant line fit to the data points in the range of $3.3$~GeV $< p_t^{\Lambda}< 4.5$~GeV yields:
$P_{\Lambda} = (-23.3 \pm 11.2) \times 10^{-2}$ with $\chi^2/ndf = 0.22/2$
for Au+Au collisions at $\sqrt{s_{NN}}$=200~GeV
(centrality region \mbox{20-70\%}) and
$P_{\Lambda} = (-20.7  \pm 14.2) \times 10^{-2}$ with $\chi^2/ndf = 0.38/2$
for Au+Au collisions at $\sqrt{s_{NN}}$=62.4~GeV
(centrality region \mbox{0-80\%}).
Unfortunately, at present there exists no theory prediction for the
global polarization dependence on particle transverse momentum to compare with these
results.

It was found in this analysis that the event plane vectors defined with the particles
measured in the Forward
TPCs are reliable within the centrality region \mbox{0-80\%} for Au+Au
collisions at $\sqrt{s_{NN}}$=62.4~GeV.
With higher multiplicity at  $\sqrt{s_{NN}}$=200~GeV,
saturation effects in the Forward TPCs
for the most central collisions become evident,
and the estimated reaction plane angle is unreliable.
Due to this effect, the centrality region used for the
$\Lambda$ ($\bar\Lambda$) hyperon global polarization measurement  in Au+Au collisions
at $\sqrt{s_{NN}}$=200~GeV is limited to \mbox{20-70\%}.

\begin{figure}[th]
\includegraphics[width=0.5\textwidth]{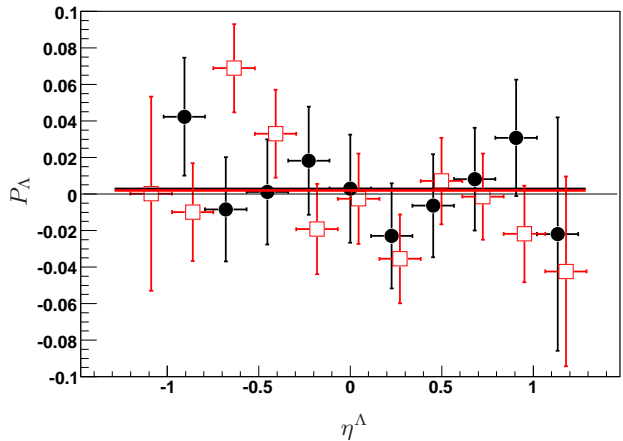}
\put(-256,82){\rotatebox{90}{$P_{\Lambda}$}} \put(-131,-5){$\eta^{\Lambda}$}
\caption{
\label{lambdaGlobalPolarization_eta} (Color online) Global polarization of
$\Lambda$--hyperons as a function of $\Lambda$ pseudorapidity $\eta^\Lambda$. Filled
circles show the results for Au+Au collisions at $\sqrt{s_{NN}}$=200~GeV (centrality
region \mbox{20-70\%}). A constant line fit to these data points yields $P_{\Lambda} = (2.8 \pm
9.6) \times 10^{-3}$ with $\chi^2/ndf = 6.5/10$. Open squares show the results for Au+Au collisions at
$\sqrt{s_{NN}}$=62.4~GeV (centrality region \mbox{0-80\%}). A constant line fit gives $P_{\Lambda}
= (1.9 \pm 8.0) \times 10^{-3}$ with $\chi^2/ndf = 14.3/10$.
Only statistical uncertainties are shown.
}
\end{figure}

Figure~\ref{lambdaGlobalPolarization_eta} presents the $\Lambda$--hyperon global
polarization as a function of $\Lambda$  pseudorapidity $\eta^{\Lambda}$. The symbol keys
for the data points are the same as in Fig.~\ref{lambdaGlobalPolarization_pt}. Note that
the scale is different from the one in Fig.~\ref{lambdaGlobalPolarization_pt}. The
$p_t$-integrated global polarization result is dominated by the region
$p^{\Lambda}_t<3$~GeV/c, where the measurements are consistent with zero
(see Fig.~\ref{lambdaGlobalPolarization_pt}). The solid line in
Fig.~\ref{lambdaGlobalPolarization_eta} indicates a constant fit to the experimental data:
$P_{\Lambda} = (2.8 \pm 9.6) \times 10^{-3}$ with $\chi^2/ndf = 6.5/10$
for Au+Au collisions at $\sqrt{s_{NN}}$=200~GeV
(centrality region \mbox{20-70\%}) and
$P_{\Lambda} = ( 1.9 \pm 8.0) \times 10^{-3}$ with $\chi^2/ndf = 14.3/10$
for Au+Au collisions at $\sqrt{s_{NN}}$=62.4~GeV
(centrality region \mbox{0-80\%}).
As indicated by the numerical values given in the caption, the lines associated with each
of the two beam energies are almost indistinguishable from zero within the resolution of
the plot. The results for the $\Lambda$--hyperon global polarization  as a function of
$\eta^\Lambda$ within the STAR acceptance are consistent with zero.

\begin{figure}[th]
\includegraphics[width=0.5\textwidth]{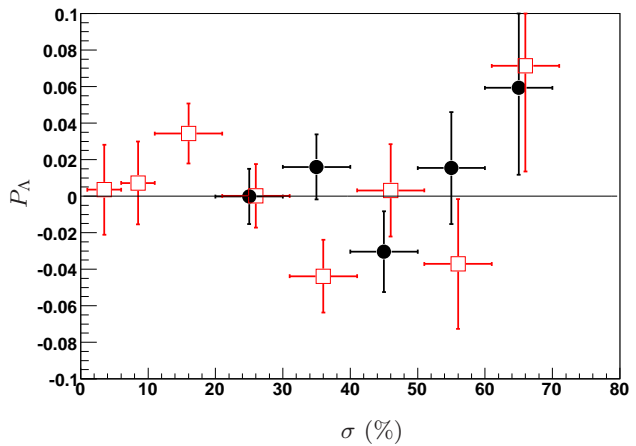}
\put(-256,82){\rotatebox{90}{$P_{\Lambda}$}}
\put(-131,-5){$\sigma$ (\%)}
\caption{
\label{lambdaGlobalPolarization_sigma}
(Color online) Global polarization of $\Lambda$--hyperons as a function of centrality
given as fraction of the total inelastic hadronic cross section.
Filled circles show the results for Au+Au collisions
at $\sqrt{s_{NN}}$=200~GeV (centrality region \mbox{20-70\%}) and
open squares indicate the results for Au+Au collisions at
$\sqrt{s_{NN}}$=62.4~GeV (centrality region \mbox{0-80\%}).
Only statistical uncertainties are shown.
}
\end{figure}

Figure~\ref{lambdaGlobalPolarization_sigma} presents the $\Lambda$--hyperon global
polarization as a function of centrality given as a fraction of the total inelastic
hadronic cross section. The symbol keys for the data points are the same as in
Fig.~\ref{lambdaGlobalPolarization_pt}. Within the statistical uncertainties we observe no
centrality-dependence of the $\Lambda$ global polarization.

\begin{figure}[th]
\includegraphics[width=0.5\textwidth]{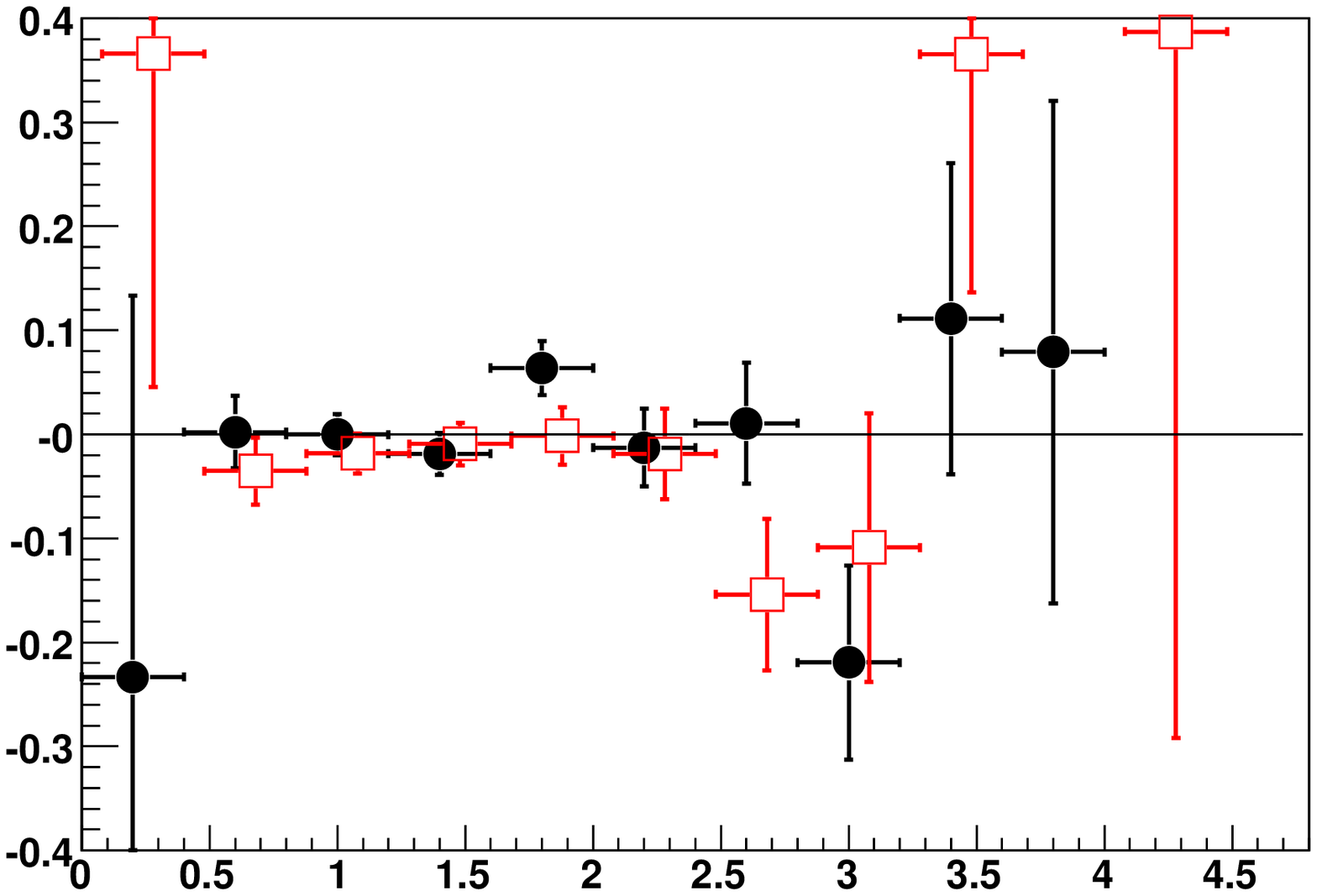}
\put(-256,82){\rotatebox{90}{$P_{\bar\Lambda}$}}
\put(-131,-5){$p_t^{\bar\Lambda}$ (GeV/c)}
\caption{\label{antiLambdaGlobalPolarization_pt}
(Color online) Global polarization of $\bar\Lambda$--hyperons as a function of $\bar\Lambda$
transverse momentum $p_t^{\bar\Lambda}$.
Filled circles show the results for Au+Au collisions
at $\sqrt{s_{NN}}$=200~GeV (centrality region \mbox{20-70\%}) and
open squares indicate the results for Au+Au collisions at
$\sqrt{s_{NN}}$=62.4~GeV (centrality region \mbox{0-80\%}).
Only statistical uncertainties are shown.
}
\end{figure}
\begin{figure}[th]
\includegraphics[width=0.5\textwidth]{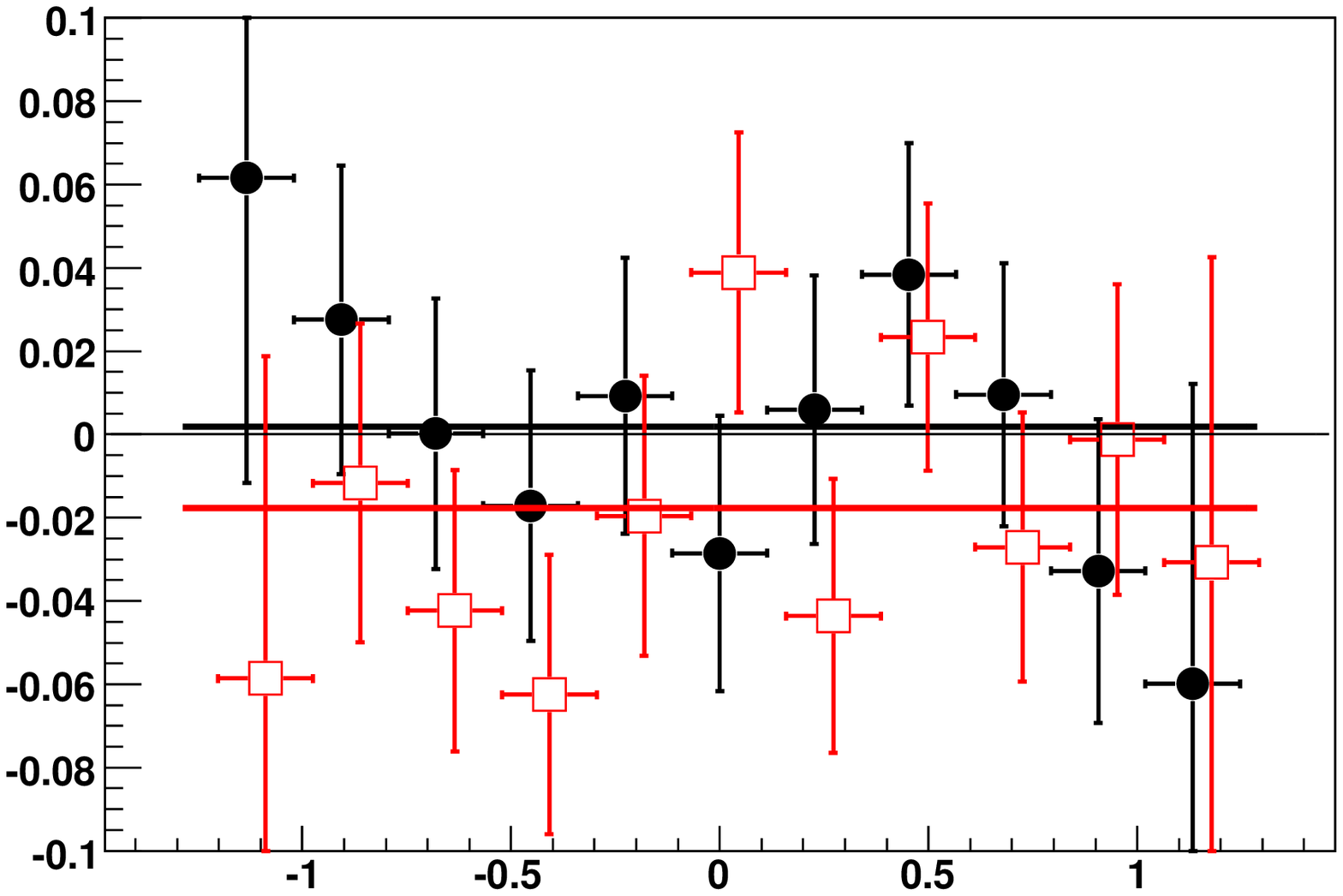}
\put(-256,82){\rotatebox{90}{$P_{\bar\Lambda}$}} \put(-131,-5){$\eta^{\bar\Lambda}$}
\caption{\label{antiLambdaGlobalPolarization_eta} (Color online) Global polarization of
$\bar\Lambda$--hyperons as a function of $\bar\Lambda$ pseudorapidity $\eta^{\bar\Lambda}$.
Filled circles show the results for Au+Au collisions at
$\sqrt{s_{NN}}$=200~GeV (centrality region \mbox{20-70\%}).
A constant line fit to these data points yields
$P_{\bar\Lambda} = (1.8 \pm 10.8) \times 10^{-3}$ with $\chi^2/ndf = 5.5/10$.
Open squares show the results for Au+Au collisions at $\sqrt{s_{NN}}$=62.4~GeV
(centrality region \mbox{0-80\%}).
A constant line fit gives
$P_{\bar\Lambda} = (-17.6 \pm 11.1) \times 10^{-3}$ with $\chi^2/ndf = 8.0/10$.
Only statistical uncertainties are shown.
}
\end{figure}
\begin{figure}[th]
\includegraphics[width=0.5\textwidth]{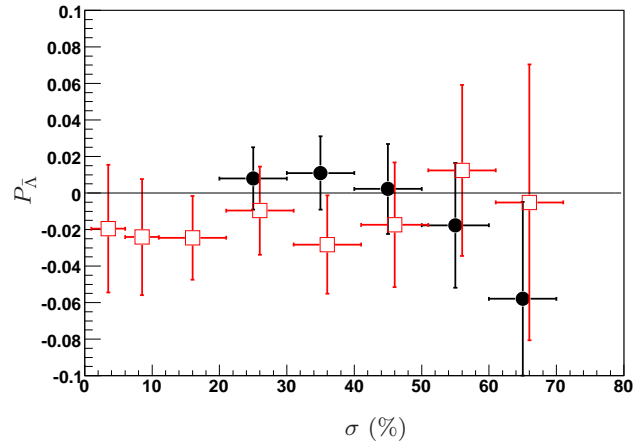}
\put(-256,82){\rotatebox{90}{$P_{\bar\Lambda}$}}
\put(-131,-5){$\sigma$ (\%)}
\caption{\label{antiLambdaGlobalPolarization_sigma}
(Color online) Global polarization of $\bar\Lambda$--hyperons as a function of centrality.
Filled circles show the results for Au+Au collisions
at $\sqrt{s_{NN}}$=200~GeV (centrality region \mbox{20-70\%}) and
open squares indicate the results for Au+Au collisions at
$\sqrt{s_{NN}}$=62.4~GeV (centrality region \mbox{0-80\%}).
Only statistical uncertainties are shown.
}
\end{figure}

The statistics for $\bar\Lambda$--hyperons are smaller than those for $\Lambda$--hyperons
by 40\% (20\%) for Au+Au collisions at $\sqrt{s_{NN}}$=62.4~GeV (200~GeV).
Figures~\ref{antiLambdaGlobalPolarization_pt}, \ref{antiLambdaGlobalPolarization_eta}, and
\ref{antiLambdaGlobalPolarization_sigma} show the results for the $\bar\Lambda$-hyperon
global polarization as a function of $\bar\Lambda$ transverse momentum, pseudorapidity,
and centrality (the symbol keys for the data points are the same as in
Figs.~\ref{lambdaGlobalPolarization_pt}, \ref{lambdaGlobalPolarization_eta}, and
\ref{lambdaGlobalPolarization_sigma}). Again, no deviation from zero has been observed
within statistical errors. The constant line fits for the $\bar\Lambda$--hyperon global
polarization give: $P_{\bar\Lambda} = (1.8 \pm 10.8) \times 10^{-3}$ with $\chi^2/ndf = 5.5/10$ for Au+Au collisions
at $\sqrt{s_{NN}}$=200~GeV (centrality region \mbox{20-70\%}) and $P_{\bar\Lambda} = (-17.6 \pm
11.1) \times 10^{-3}$ with $\chi^2/ndf = 8.0/10$ for Au+Au collisions at $\sqrt{s_{NN}}$=62.4~GeV (centrality region
\mbox{0-80\%}).

\subsection{Acceptance effects and systematic uncertainties}
\label{SystematicsUncertainties}

The derivation of  Eq.~\ref{GlobalPolarizationObservable} assumes a perfect reconstruction
acceptance for hyperons. For the case of an imperfect detector, we similarly consider the
average of $\langle \sin \left( \phi^*_p - \Psi_{\rm RP}\right)\rangle$ but take into account
the fact that the integral over solid angle $d\Omega^*_p = d\phi^*_p \sin \theta^*_p d
\theta^*_p$ of the hyperon decay baryon's 3-momentum ${\bf p}^*_p$ in the hyperon rest
frame is affected by detector acceptance:
\begin{widetext}
\begin{eqnarray}
\label{meanSinAcc}
{\langle \sin \left( \phi^*_p - \Psi_{\rm RP}\right)\rangle =}
\int {\frac{d\Omega^*_p}{4\pi}\frac{d\phi_H}{2\pi}
 A({\bf p}_H, {\bf p}^*_p) \int\limits_0^{2\pi} \frac{d\Psi_{\rm RP}}{2\pi} \sin \left( \phi^*_p - \Psi_{\rm RP}\right)}
\left[1+\alpha_H~P_H ({\bf p}_H; \Psi_{\rm RP})~ \sin \theta^*_p \cdot \sin \left( \phi^*_p - \Psi_{\rm RP}\right)\right].
\end{eqnarray}
Here ${\bf p}_H$ is the hyperon 3-momentum, and $A\left({\bf p}_H, {\bf p}^*_p\right)$ is
a function to account for detector acceptance.
The integral of this function over $(d\Omega^*_p/4\pi)(d\phi_H/2\pi)$ is normalized to unity.
As stated in the beginning of section~\ref{GlobalPolarization},
the global polarization can depend on the relative azimuthal
angle ($\phi_H-\Psi_{\rm RP}$). Taking into account the symmetry of the system, one can expand
the global polarization as a function of ($\phi_H-\Psi_{\rm RP}$) in a sum over even
harmonics:
\begin{eqnarray}
\label{sumForGlobalPolarization}
P_H\left(\phi_H-\Psi_{\rm RP},p_t^H,\eta^H\right)=\sum_{n=0}^\infty P_H^{(2n)}\left(p_t^H,\eta^H\right)\cos\{2n[\phi_H-\Psi_{\rm RP}]\}.
\end{eqnarray}
In this work we report the global polarization averaged over all possible values of ($\phi_H-\Psi_{\rm RP}$):
\begin{eqnarray}
P_H\left(p_t^H,\eta^H\right) \equiv \overline{P_H\left(\phi_H-\Psi_{\rm RP},p_t^H,\eta^H\right)} = P_H^{(0)}\left(p_t^H,\eta^H\right).
\end{eqnarray}
Substitution of Eq.~\ref{sumForGlobalPolarization} into Eq.~\ref{meanSinAcc}
and integration over the reaction plane angle $\Psi_{\rm RP}$ gives:
\begin{eqnarray}
\label{GlobalPolarizationSinAcc}
\langle \sin \left( \phi^*_p - \Psi_{\rm RP}\right)\rangle =
\frac{\alpha_H}{2} \int {\frac{d\Omega^*_p}{4\pi} \frac{d\phi_H}{2\pi}A\left({\bf p}_H, {\bf p}^*_p\right) \sin\theta^*_p
\left[P_H\left(p_t^H,\eta^H\right) -\frac{1}{2}\cos\left[2(\phi_H-\phi_p^*)\right]~P_H^{(2)}\left(p_t^H,\eta^H\right) \right]}.
\end{eqnarray}
Based on this equation, the observable (\ref{GlobalPolarizationObservable}) can be written in the following form:
\begin{eqnarray}
\nonumber
\frac{8}{\pi\alpha_H}\langle \sin \left( \phi^*_p - \Psi_{\rm RP}\right)\rangle &=&
\frac{4}{\pi}~\overline{\sin\theta^*_p}~P_H\left(p_t^H,\eta^H\right)
- \frac{2}{\pi}~\overline{\sin\theta^*_p\cos\left[2(\phi_H-\phi_p^*)\right]}~P_H^{(2)}\left(p_t^H,\eta^H\right)\\
\label{GlobalPolarizationObservableAcc}
& = & A_{0}(p_t^H,\eta^H)~P_H\left(p_t^H,\eta^H\right) - A_{2}(p_t^H,\eta^H)~P_H^{(2)}(p_t^H,\eta^H),
\end{eqnarray}
where functions $A_{0}(p_t^H,\eta^H)$ and $A_{2}(p_t^H,\eta^H)$ are defined by the average of
$\sin\theta^*_p$ and $\sin\theta^*_p\cos\left[2(\phi_H-\phi_p^*)\right]$ over detector acceptance according to equations:
\begin{eqnarray}
\label{AccCoefficient}
A_{0}(p_t^H,\eta^H) = & {\displaystyle \frac{4}{\pi}}~\overline{\sin\theta^*_p} & \equiv \frac{4}{\pi} \int {\frac{d\Omega^*_p}{4\pi} \frac{d\phi_H}{2\pi}A\left({\bf p}_H, {\bf p}^*_p\right) \sin\theta^*_p}.
\\
\label{AccCoefficientAdditive}
A_{2}(p_t^H,\eta^H) = &{\displaystyle \frac{2}{\pi}}~\overline{\sin\theta^*_p\cos\left[2(\phi_H-\phi_p^*)\right]}
& \equiv \frac{2}{\pi} \int {\frac{d\Omega^*_p}{4\pi} \frac{d\phi_H}{2\pi}A\left({\bf p}_H, {\bf p}^*_p\right) \sin\theta^*_p\cos\left[2(\phi_H-\phi_p^*)\right]}.
\end{eqnarray}
\end{widetext}
As follows from Eq.~\ref{GlobalPolarizationObservableAcc}
there exist two different contributions due to detector acceptance.
The first one affects the overall scale of the measured global polarization
and is given by the acceptance correction function $A_{0}(p_t^H,\eta^H)$.
A different effect due to non-uniform detector acceptance
comes from the admixture of higher harmonic terms in Eq.~\ref{GlobalPolarizationObservableAcc}
proportional to $P_H^{(2)}\left(p_t^H,\eta^H\right)$.
Since the values of $P_H^{(2)}\left(p_t^H,\eta^H\right)$ are not measured in this analysis
and values of $A_{2}(p_t^H,\eta^H)$ are small (see below),
we present data in Figs.~\ref{lambdaGlobalPolarization_pt}-\ref{antiLambdaGlobalPolarization_sigma}
corrected only with the $A_{0}(p_t^H,\eta^H)$ function,
providing an estimate for the systematic uncertainty associated with acceptance effects due to higher harmonic terms.
In the case of a perfect acceptance, \mbox{$A_0(p_t^H,\eta^H)=1$} and \mbox{$A_2(p_t^H,\eta^H)=0$},
and Eq.~\ref{GlobalPolarizationObservableAcc} reduces to the global polarization~(\ref{GlobalPolarizationObservable}).
Since the background contribution to the hyperon invariant mass distribution is small (see
Fig.~\ref{lambdaInvMass}), the value of these functions $A_0(p_t^H,\eta^H)$ and $A_2(p_t^H,\eta^H)$ can be extracted
directly from the experimental data by calculating the average
over all events for the $\Lambda$ and $\bar\Lambda$ candidates for
$\sin\theta^*_p$ and $\sin\theta^*_p\cos\left[2(\phi_H-\phi_p^*)\right]$, respectively.

\begin{figure}[th]
\includegraphics[width=0.5\textwidth]{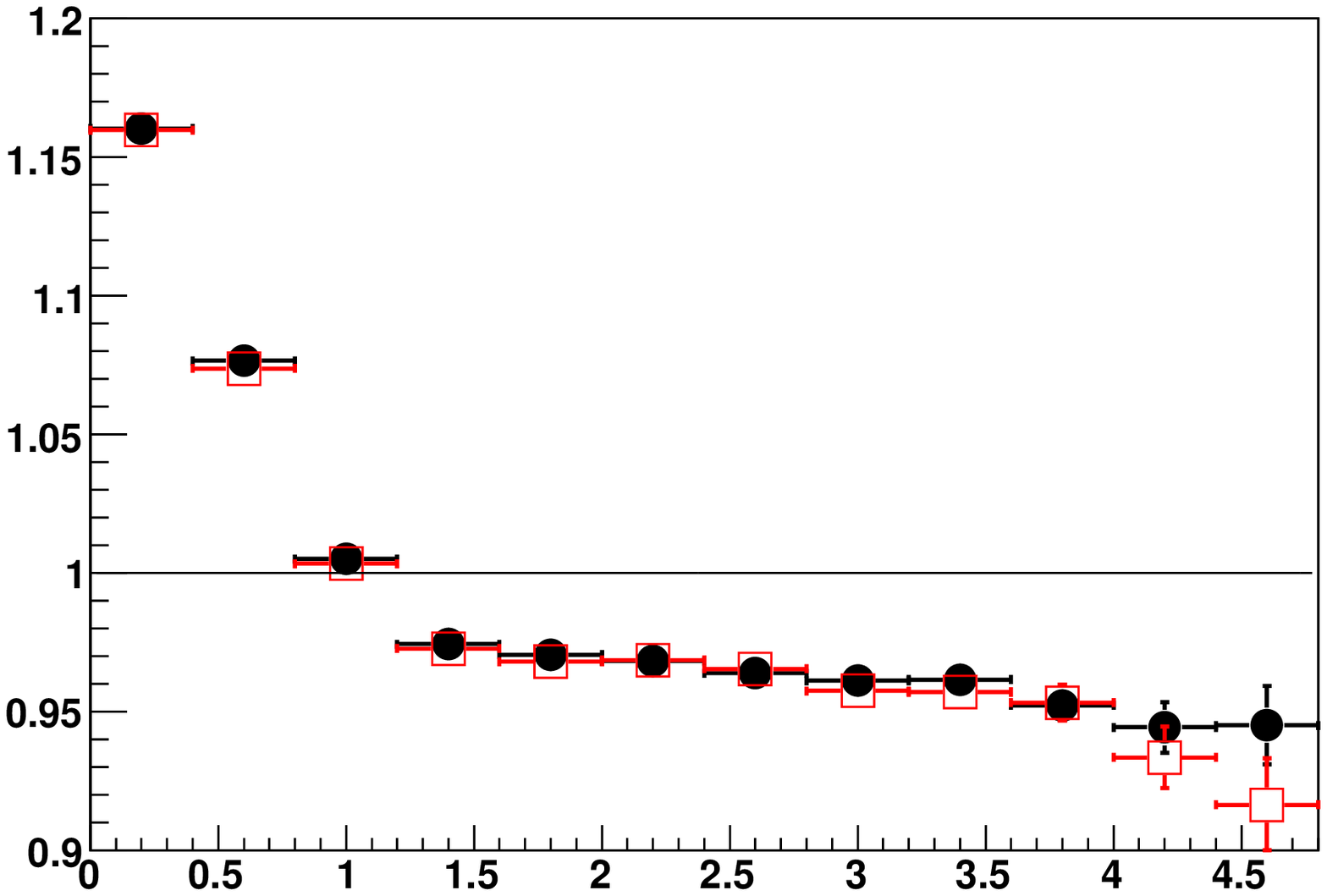}
\put(-260,70){\rotatebox{90}{$A_{0}^{\Lambda,\bar\Lambda}$}}
\put(-130,-5){$p_t^{\Lambda,\bar\Lambda}$ (GeV/c)}\\
\includegraphics[width=0.5\textwidth]{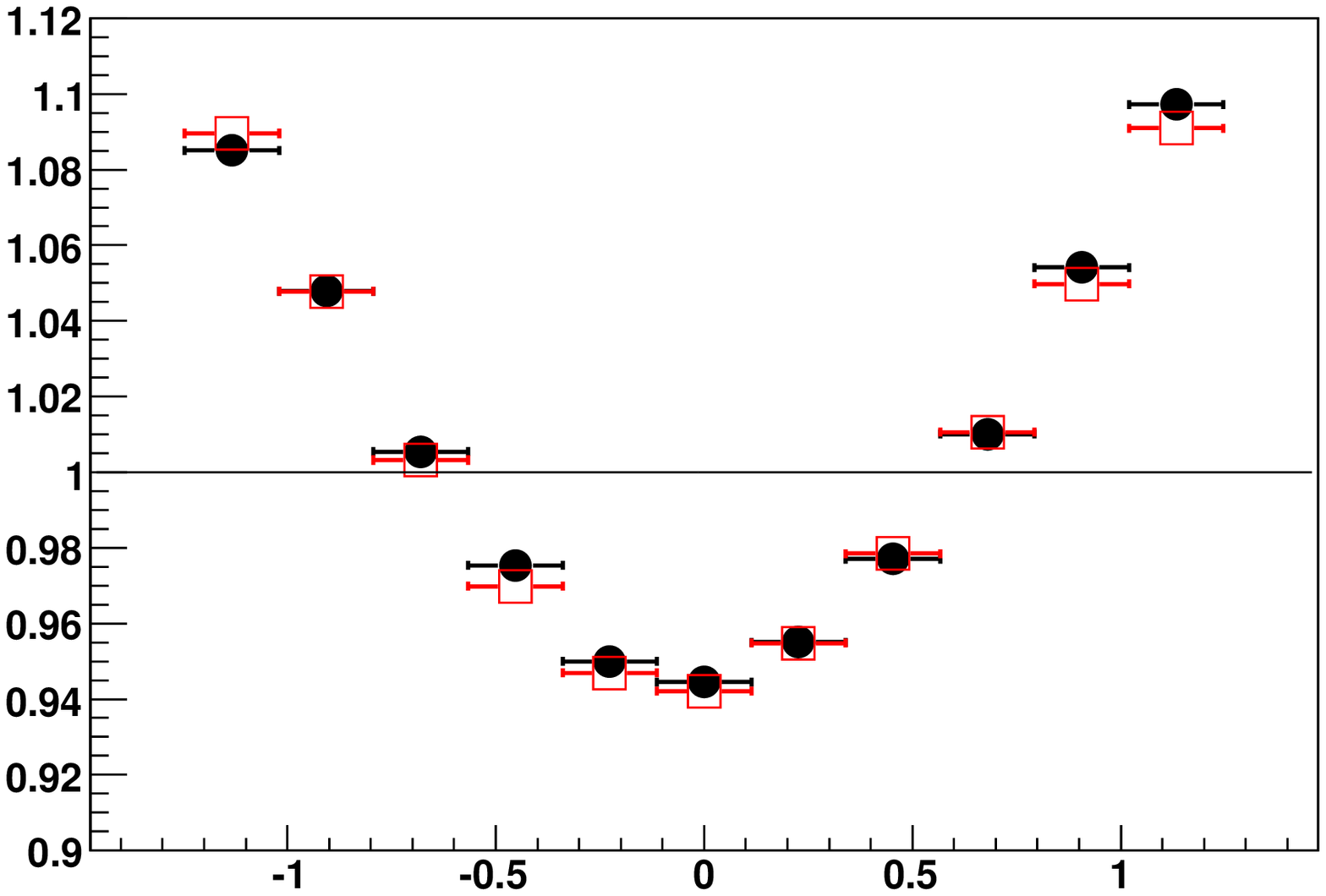}
\put(-260,70){\rotatebox{90}{$A_{0}^{\Lambda,\bar\Lambda}$}}
\put(-130,-5){$\eta^{\Lambda,\bar\Lambda}$} \caption{ \label{corrFigure} (Color online)
Acceptance correction function $A_0(p_t^H,\eta^H)$ defined in (\ref{AccCoefficient})
as a function of $\Lambda$ (filled circles) and
$\bar\Lambda$ (open squares) transverse momentum (top) and pseudorapidity (bottom).
The deviation of this function from unity affects  the overall scale of
the measured global polarization according to Eq.~\ref{GlobalPolarizationObservableAcc}.
See the text for details and discussions on
$A_0$ $p_t^H$ and $\eta^H$ dependence.}
\end{figure}

Figure~\ref{corrFigure} shows the pseudorapidity $\eta^{\Lambda,\bar\Lambda}$ and
transverse momentum $p_t^{\Lambda,\bar\Lambda}$ dependence
of the acceptance correction function $A_{0}$ defined in Eq.~\ref{AccCoefficient} for $\Lambda$ (filled circles) and
$\bar\Lambda$ (open squares) candidates reconstructed from Au+Au collisions at
$\sqrt{s_{NN}}$=200~GeV.
For different centralities, this function varies within 2\% around an average value of 0.98.
The deviation of $A_{0}$ from unity is small
and it reflects losses of the daughter protons (anti-protons) or pions
from the STAR detector acceptance,
primarily at small angles with respect to the beam direction.
Proton (anti-proton) losses and pion losses dominate in different regions of phase space,
since in the detector frame the protons (anti-protons) follow the parent $\Lambda$ ($\bar\Lambda$)
direction much more closely than do the pions.  When the $\Lambda$ ($\bar\Lambda$)
momentum is itself near the acceptance edges ($|\eta| \approx 1$),
then the primary losses come from protons (anti-protons) falling even closer to the beam direction.
This disfavoring of small $\theta_p^*$ tends to
increase $\overline{\sin \theta_p^*}$, hence $A_0$, with respect to uniform acceptance.
In contrast, when the hyperon is near midrapidity or at high $p_t^H$,
the daughter protons are constrained
to stay within the detector acceptance.
Then the primary losses arise from forward-going daughter pions,
preferentially correlated with large $\sin \theta_p^*$,
tending to reduce $A_0$ from unity.
In any case, the corresponding corrections to the absolute value
of the global polarization are estimated to be less than 20\% of the
extracted polarization values.

\begin{figure}[th]
\includegraphics[width=0.5\textwidth]{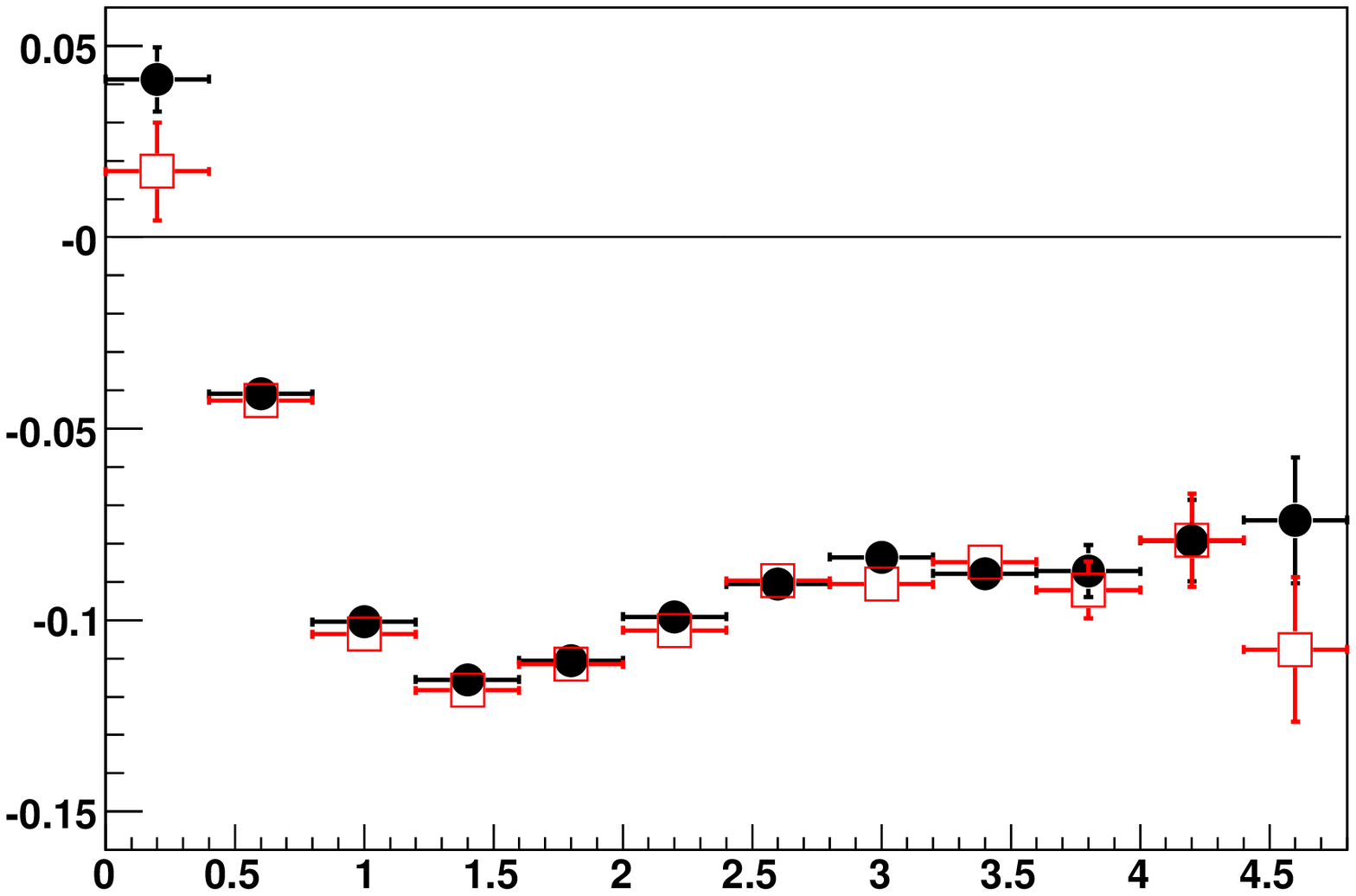}
\put(-260,70){\rotatebox{90}{$A_{2}^{\Lambda,\bar\Lambda}$}}
\put(-130,-5){$p_t^{\Lambda,\bar\Lambda}$ (GeV/c)}\\
\includegraphics[width=0.5\textwidth]{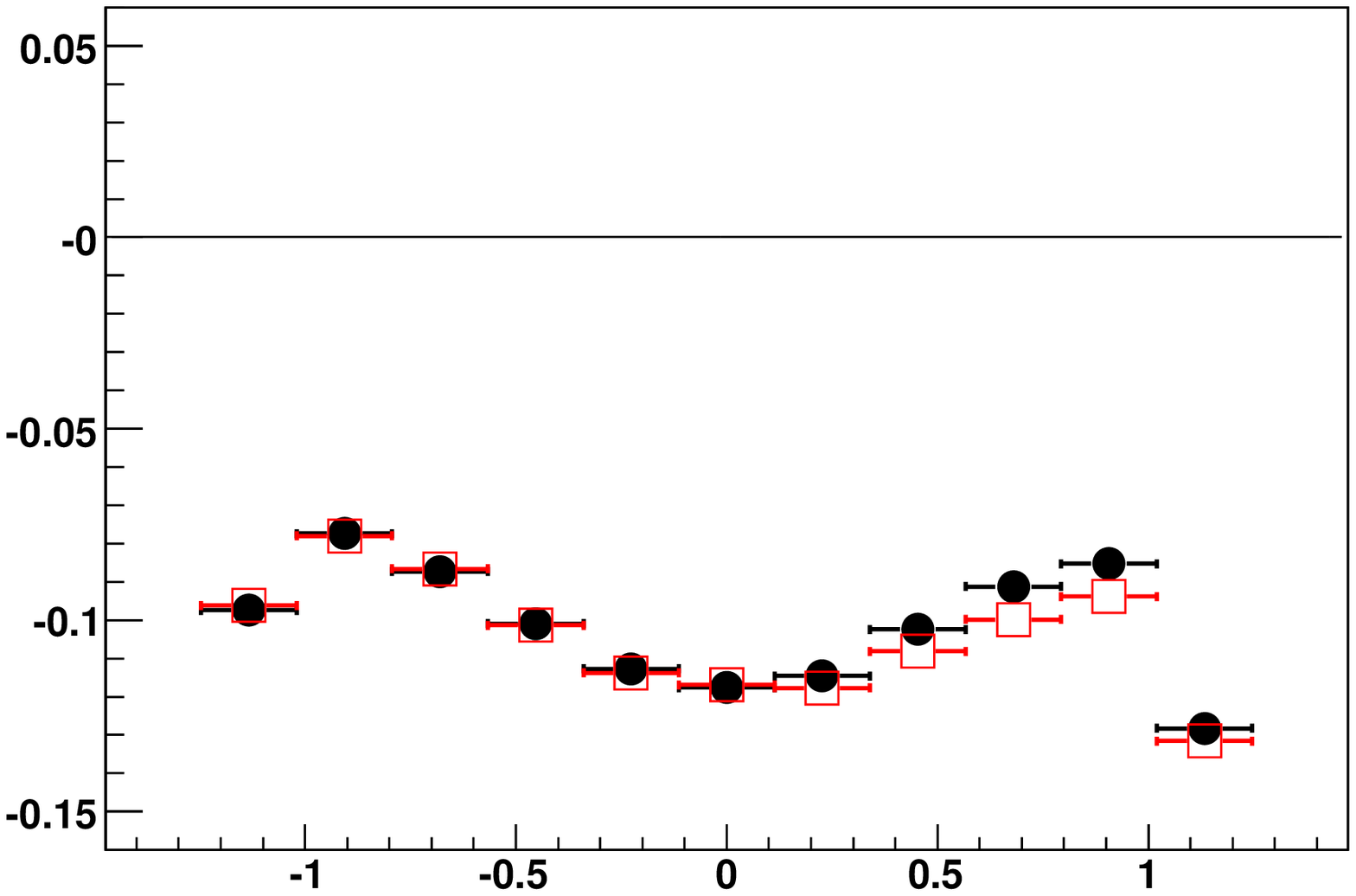}
\put(-260,70){\rotatebox{90}{$A_{2}^{\Lambda,\bar\Lambda}$}}
\put(-130,-5){$\eta^{\Lambda,\bar\Lambda}$} \caption{ \label{corr2Figure} (Color online)
Function $A_2(p_t^H,\eta^H)$ defined in (\ref{AccCoefficientAdditive}) as a function of $\Lambda$ (filled circles) and
$\bar\Lambda$ (open squares) transverse momentum (top) and pseudorapidity (bottom).
The deviation of this function from zero defines the contribution to the observable (\ref{GlobalPolarizationObservable}) from
$P_H^{(2)}\left(p_t^H,\eta^H\right)$ in the expansion (\ref{sumForGlobalPolarization}).}
\end{figure}

The contribution from $P_H^{(2)}\left(p_t^H,\eta^H\right)$ in Eq.~\ref{GlobalPolarizationObservableAcc}
is defined by the deviation from zero of the function $A_2(p_t^H,\eta^H)$.
The value of this function has been also extracted from the experimental data and is presented in
Fig.~\ref{corr2Figure}.
The global polarization $P_H$ is not expected to change
sign depending on the relative orientation
of the hyperons momentum direction and the system orbital momentum.
This implies that $|P_H^{(2)}| \lesssim |P_H^{(0)}|$, and
the corresponding corrections from the admixture of $P_H^{(2)}\left(p_t^H,\eta^H\right)$ to the
$\Lambda$ and $\bar\Lambda$ hyperon global polarization measurement
are less than $A_2$, which is $<$20\%.

The hyperon directed flow is defined as the first order coefficient
in the Fourier expansion of the hyperon azimuthal angular distribution with respect to the reaction plane.
Due to non-uniform detector acceptance it will interfere with the hyperon global polarization measurement
and this can dilute the measured polarization \cite{Selyuzhenkov:2006tj}.
Assuming that hyperon directed flow is of the same order of magnitude as for charged particles ($\le 10$\%),
the effect of such interference is negligible ($\le 1$\%)
in the $\Lambda$ and $\bar\Lambda$ hyperon global polarization measurement \cite{Selyuzhenkov:2006tj}.
It is possible that due to both the hyperon reconstruction
procedure and imperfection of the reaction plane determination,
the higher harmonics of hyperon anisotropic flow (i.e. elliptic flow) will also contribute,
but these are higher order corrections compared to those from hyperon directed flow.

To check the analysis code, Monte Carlo simulations
with sizable linear transverse momentum dependence
of hyperon global polarization and hydrodynamic $p_t^H$
spectra have been performed.
Both the sign and magnitude of the reconstructed
polarization agreed with the input values within statistical uncertainties.

The measurement could be affected by other systematic effects.
Most of them are similar to those present in an anisotropic
flow analysis, with the most significant one coming from the determination of the event
plane vector and its resolution. In calculating the reaction plane resolution, we have
used the random sub-event technique \cite{Poskanzer:1998yz}, as well as the mixed harmonic
method \cite{Poskanzer:1998yz,Adams:2004bi,Adams:2005ca} with the second-order event plane
determined from TPC tracks. The mixed harmonic method is known to be effective in
suppressing a wide range of non-flow effects (short range correlations, effects of
momentum conservation \cite{Borghini:2002mv}, etc.).

To suppress the contribution to the global polarization measurement from ``non-flow"
effects (mainly due to momentum conservation) the combination of both east and west
Forward TPC event plane vectors was used. The contribution from other few-particle
correlations (i.e., resonances, jets, etc.) was estimated by comparing the results
obtained from correlations using positive or negative particles to determine the reaction
plane. Uncertainties related to the dependence of tracking efficiency (in particular,
charged particle and $\Lambda$ ($\bar\Lambda$) hyperon reconstruction efficiency) on azimuthal angle were
estimated by comparing the results obtained with different magnetic field settings and
also with event plane vectors determined from positively or negatively charged particles.
The magnitude of non-flow correlations is multiplicity dependent and its contribution to
anisotropic flow measurement increases with collision centrality.
The average uncertainty due to the reaction plane reconstruction is estimated to be 30\%.

\begin{table}[th]
\begin{tabular}{ll}
\hline
Source of uncertainty & value\\
\hline
Decay parameter $\alpha_{\Lambda,\bar\Lambda}$ error& 2\%\\
Background, $K^0_S$ contamination & 8\% \\
Multistrange feed-down & 15\% \\
$\Sigma^0$ feed-down & 30\% \\
$P_H(\phi_H - \psi_{\rm RP})$ dependence ($A_2$ term)& 20\% \\
Reaction plane uncertainty& 30\% \\
Hyperon anisotropic flow contribution& $\le 1$\%\\
Hyperon spin precession&  $\le 0.1$\%\\
\hline
Total uncertainty (sum) & 105\%\\
\end{tabular}
\caption{\label{systematicSummaryTable}
Summary table for systematic uncertainties of the $\Lambda$ ($\bar\Lambda$) global polarization measurement.
See sections~\ref{Technique} and \ref{SystematicsUncertainties} for details.}
\end{table}

All uncertainties discussed in sections~\ref{Technique} and \ref{SystematicsUncertainties} are relative.
Table~\ref{systematicSummaryTable} summarizes systematic errors in the global
polarization measurement.
Although some of the systematic uncertainty contributions may be
expected to be correlated, we have conservatively combined all
contributions by linear summation to arrive at an upper limit for the
total systematic uncertainty.
The overall relative uncertainty in the $\Lambda$
($\bar\Lambda$) hyperon global polarization measurement due to detector effects is
estimated to be less than a factor of 2.

Taking all these possible correction factors into account,
and that our measurements are consistent with zero with statistical error of about $0.01$,
our results suggest that the global $\Lambda$ and $\bar\Lambda$
polarizations are $\leq 0.02$ in magnitude.
\section{Conclusion}
\label{Conclusion}

The $\Lambda$ and $\bar\Lambda$ hyperon global polarization has been measured in Au+Au
collisions at center of mass energies $\sqrt{s_{NN}}$=62.4 and 200~GeV with the STAR
detector at RHIC. An upper limit of $|P_{\Lambda,\bar\Lambda}| \leq 0.02$ for the global
polarization of $\Lambda$ and $\bar\Lambda$ hyperons within the STAR acceptance is
obtained. This upper limit is far below the few tens of percent values discussed
in \cite{Liang:2004ph}, but it falls within the predicted region from the more realistic
calculations \cite{Liang:2007ma} based on the HTL (Hard Thermal Loop) model.

\section*{Acknowledgments}
We thank the RHIC Operations Group and RCF at BNL, and the
NERSC Center at LBNL for their support. This work was supported
in part by the Offices of NP and HEP within the U.S. DOE Office 
of Science; the U.S. NSF; the BMBF of Germany; CNRS/IN2P3, RA, RPL, and
EMN of France; EPSRC of the United Kingdom; FAPESP of Brazil;
the Russian Ministry of Science and Technology; the Ministry of
Education and the NNSFC of China; IRP and GA of the Czech Republic,
FOM of the Netherlands, DAE, DST, and CSIR of the Government
of India; Swiss NSF; the Polish State Committee for Scientific 
Research; SRDA of Slovakia, and the Korea Sci. \& Eng. Foundation.

\bibliography{globalPolarization}

\end{document}